# Detailed magneto heat capacity analysis of SnAs topological superconductor

M. M. Sharma[1,2,#] & V.P.S. Awana[1,2,*]

[1]*National Physical Laboratory (CSIR), Dr. K. S. Krishnan Road, New Delhi 110012, India*

[2]*Academy of Scientific and Innovative Research, Ghaziabad, U.P. 201002, India*

**Abstract:**

In this article, we report magneto heat capacity analysis of superconducting SnAs. Magneto heat capacity analysis of superconductors is an important tool to determine bulk superconductivity as well as the pairing mechanism of Cooper pairs. SnAs crystal is characterized through X-ray diffraction (XRD) and X-ray photoelectron spectroscopy (XPS). Magneto transport measurements of studied SnAs superconductor evidenced presence of superconductivity at around 4K, which persists up to an applied field of 250Oe. The bulk nature of superconductivity is determined through AC susceptibility ($\chi$) along with the heat capacity measurements. Magneto heat capacity measurements show SnAs to be a fully gapped s wave superconductor. This finding is well supported by calculated physical parameters like $\alpha$ (3.36), $\lambda_{e\text{-}ph}$ (0.70) and $\Delta C_{el}/\gamma T_c$ (1.41). Calculation of residual Sommerfeld coefficient ($\gamma_{res}$) at different fields, confirms node-less superconductivity in SnAs.

**Keywords:** Superconductivity, X-ray Photoelectron Spectroscopy, AC susceptibility, Heat Capacity, Node-less superconductivity.

#: E-mail: manish.npl18a@acsir.res.in

*Corresponding Author
Dr. V. P. S. Awana:  E-mail: awana@nplindia.org
Ph. +91-11-45609357, Fax-+91-11-45609310
Homepage: awanavps.webs.com

**Introduction:**

Ever since the discovery of superconductors, superconductivity continues to astonish the condensed matter scientists. In particular, the discovery of copper and iron-based high $T_c$ superconductors kept this field young and alive [1-3]. More recently, some superconductors were found to show topological effects in their normal state [4-6]. This class of superconductors is known as topological superconductors. Topological superconductors show fully gapped superconductivity below $T_c$ and the domination of topological surface states above [7,8]. This unique combination of higher temperature topology and low temperature superconductivity enables superconductors to host Majorana Fermions [8]. Majorana fermions are the fermionic particles, which are their own antiparticles, and realization of these particles is crucial in the field of fault tolerant quantum computing [8,9]. Some of the doped TIs are found to show topological superconductivity with enhancement in the charge carrier density [10-14]. There are very few materials that show topological superconductivity in intrinsic form. Strong spin-orbit coupling (SOC) is the key parameter to realize topological superconductivity. Compounds with elements having higher atomic numbers, show strong SOC; this makes the study of superconductors containing heavier elements crucial to observe topological superconductivity. Some of the Sn based superconductors e.g. $Sn_{1-x}Sb_x$, $In_xSn_{1-x}Te$ and SnAs are studied recently in the context to topological superconductivity due to presence of strong SOC [15-19]. Apart from strong SOC in SnAs, the calculated Z2 invariants using first principle calculations do suggest the presence of strong topology as well in the same material [19].

The superconducting properties of SnAs are though known from a long time [20], but its type of superconductivity is yet inconclusive [18-23]. Though SnAs is reported to be type-I superconductor in ref. [21], yet there are clear indications towards weak type-II superconductivity in magnetization measurements in ref. [18,19]. A similar ambiguity about the type of superconductivity can be found in $PdTe_2$, which was shown to be a type-I [5] as well type-II or mixed type-I and type-II superconductor [24-26]. Magnetic field dependence of heat capacity jump helps to determine the nature of superconducting gap, which is discussed in ref. 18. There exist only two reports on heat capacity measurements of SnAs [18,21]. The parameters obtained from heat capacity measurements in ref. 18 and ref. 21 contradict each other. In ref. 21, SnAs is reported to be within BCS weak coupling limit, on the other hand the same is shown to be well above the BCS weak coupling limit in ref. 18. Also, the Scanning tunnelling spectroscopic (STS) measurements suggested SnAs to be a conventional superconductor with isotropic gap [18,22]. The order parameter obtained from STS measurements in both available reports [18,22], indicated that the superconductivity of SnAs is well defined under BCS weak coupling limit. The contradiction between the two available reports on heat capacity measurements [18,21], motivates us to study more about the nature of bulk superconductivity of SnAs, through detailed analysis of magneto heat capacity measurements.

In this article, we report magneto heat capacity analysis of SnAs. Phase purity and surface chemistry of SnAs are analysed through XRD and XPS measurements. Magneto transport and AC susceptibility measurements show the presence of superconductivity around 4K. The parameters obtained from magneto heat capacity measurements suggest SnAs to be a conventional BCS superconductor with isotropic superconducting gap. The obtained value of superconducting gap at absolute zero $\Delta(0)$ in our report is in good agreement with that obtained from STS measurements in ref. 18 and ref. 22 [18,22]. Also the superconductivity is observed above 180Oe in both magneto transport and magneto heat capacity measurements as seen earlier in ref. [18]. This strengthens our claim in ref. 19 that critical field of SnAs is higher than that as reported in ref. 21. Further magneto heat capacity analysis of SnAs shows the same to be a fully gapped s wave superconductor.

**Experimental:**

The studied SnAs crystal was synthesized by following a two-step method based on solid state reaction process. Detailed synthesis process is reported in our previous report [19]. X-Ray spectra were recorded using Rigaku Miniflex II X-Ray diffractometer equipped with Cu-K$_\alpha$ radiation of 1.5418Å wavelength. XPS measurements are carried out using PHI-5000 Versa Probe III made X-ray photoelectron spectrometer. Al-K$_\alpha$ radiation is used to analyse peaks of C 1s peak and peaks of constituent elements of SnAs viz. Sn 3d & As 3d. Magneto transport measurements, AC susceptibility measurements and magneto heat capacity measurements are carried out on Quantum Design Physical Property Measurement System (QD-PPMS). Magneto transport measurements are performed by following a standard four-probe method, the current was limited to 5mA. In AC susceptibility measurements the DC background field was set to 0Oe.

**Results & Discussion:**

Phase purity and lattice parameters of synthesized crystal are determined through Rietveld refined XRD pattern, shown in Fig. 1. The refined XRD pattern shows that SnAs crystallizes in cubic structure with F m -3 m space group symmetry. The model used for Rietveld refinement treats the observed XRD peaks intensities as $y_{o,i}$ and the calculated intensities as $y_{c,i}$. Basically, Rietveld refinement optimizes the model function in a way to minimize $\Sigma_i w_i(y_{c,i} - y_{o,i})^2$. In this expression, $w_i$ represents the weight and given by $\frac{1}{\sigma^2[y_{o,i}]}$, where $\sigma[y_{o,i}]$ is standard deviation of observed intensity. The weighted profile R-factor ($R_{wp}$) is considered to be discrepancy index and given by $\frac{\Sigma_i w_i(y_{c,i}-y_{o,i})^2}{\Sigma_i w_i(y_{o,i})^2}$. The observed data is fitted to make the expected value of $w_i(y_{c,i} - y_{o,i})^2$ to be equal to 1. In this condition, the best possible value of $R_{wp}$ is termed as expected R-factor ($R_{exp}$) and given by $R_{exp}^2 = \frac{N}{\Sigma_i w_i(y_{o,i})^2}$, in this expression, N represents number of labelled data points. The values of these R-factors are very crucial to determine how well the fitting model fits the intensities, peak positions and background of observed XRD data. The parameter of goodness of fit is given by $\chi^2$ which is determined as the square of ratio of $R_{wp}$ and $R_{exp}$ i.e., $\chi^2 = \left(\frac{R_{wp}}{R_{exp}}\right)^2$. Here, the observed XRD data is well fitted with NaCl-type structure. The obtained lattice parameters along with the refinement parameters are listed in Table-1. The crystallographic information file (CIF) generated from Rietveld refinement is used to draw unit cell of SnAs using VESTA software and the same is shown in inset of Fig. 1. Sn and As atoms occupy respectively (0.5,0.5,0.5) and (0,0,0) atomic positions in cubic unit cell of SnAs.

XPS spectra is recorded for chemical analysis of synthesized SnAs crystal. SnAs has two constituent elements viz. Sn and As, hence the XPS spectra is recorded in $Sn3d^5$ and $As3d^5$ regions, which are shown in Fig. 2(a) and 2(b) respectively. The XPS peaks are calibrated with peak position of C 1s. The deconvoluted XPS spectra of SnAs in Sn $3d^5$ region is shown in Fig. 2(a). Peaks of spin orbit doublet of Sn viz. $3d_{5/2}$ and $3d_{3/2}$ are found to be at 484.20±0.01eV and 492.61±0.01eV respectively. These peaks are separated by 8.41eV, which is in complete agreement with standard value of 8.41eV as given in ref. [27]. The XPS peaks in Sn 3d region are accompanied by peaks of SnO, these peaks are generated due to surface oxidation and should not be treated as impurity in the sample, as no peak of SnO is observed in PXRD pattern. Fig. 2(b) shows deconvoluted XPS spectra in As 3d region. The spin orbit doublets of As 3d viz. $3d_{5/2}$ and $3d_{3/2}$ are merged into a single broad peak at around 40eV. This broad peak is deconvoluted into two peaks to determine the positions of As $3d_{5/2}$ and As $3d_{3/2}$, the peak position for these spin orbit doublets are found to be 39.95±0.03eV and 40.65±0.03eV respectively. The peak positions of spin orbit doublets can be verified by matching the difference of binding energy of these peaks with the standard value. Here, XPS peaks of spin orbit doublet of As 3d are found to be separated by 0.70eV, which closely matches with the standard value of 0.69eV as given in ref. [27]. XPS peaks of As 3d are also accompanied with peak of $As_2O_3$. The peaks of metal oxide in XPS spectra are occurred due to air exposure of the sample, which results in surface oxidation. All observed XPS peaks are listed in Table-2 with their respective peak positions and full width at half maxima (FWHM).

AC susceptibility measurements of SnAs are shown in Fig. 3. Bulk superconducting transition with $T_c^{onset}$ at 3.8K is evident from both real (M′) and imaginary (M′′) signals. DC magnetic field was set to 0 (with in PPMS limits) throughout the experiment, while the AC field amplitude is varied from 1-11Oe in steps of 2Oe. It is important to maintain the DC field to close to 0Oe because a slight change in DC field results in a change in $T_c$ of the superconductor. It is clear from Fig. 2 that $T_c^{onset}$ remains nearly unaltered, while changing the amplitude of AC field. It is quite common to observe a shift in $T_c^{onset}$, while changing the amplitude of AC field for a granular superconductor [28]. In our case there is nearly no change in $T_c^{onset}$ with AC amplitude, suggesting that the sample is crystalline in nature and granularity of the sample does not play a role in the superconductivity of the sample.

Fig. 4(a) depicts the transport measurement of synthesized SnAs crystal. Superconducting transition is observed at around 4K, the zoomed view of superconducting transition is shown in inset of Fig. 4(a). The normal state ρ-T plot (up to 90K) is fitted with the following equation,

$$\rho = \rho_0 + A * T^n \quad (1)$$

In above equation $\rho_0$ represents the residual resistivity and the value of n determines the nature of scattering in the sample. The value of n is found to be 2.8, suggesting phonon assisted scattering in the synthesized SnAs crystal [29]. Further, the ρ-T data is fitted in whole range using Bloch-Grüneisen formula, which is given as

$$\rho(T) = \rho_0 + \rho_{el\text{-}ph}(T) \quad (2)$$

here ρ(0) represents residual resistivity arising due to impurity scattering, the second term $\rho_{el\text{-}ph}(T)$ represents temperature dependent term, which depends on electron-phonon scattering. $\rho_{el\text{-}ph}(T)$ is given by the following formula

$$\rho_{el-ph}(T) = \alpha_{el-ph} \left(\frac{T}{\theta_D}\right)^n \int_0^{\frac{\theta_D}{T}} \frac{x^n}{(1-e^{-x})*(e^x+1)} dx \quad (3)$$

here $\alpha_{el-ph}$ is electron-phonon coupling parameter, $\theta_D$ represents Debye temperature and n is constant. ρ-T data is well fitted with the above equation for n=5, signifying dominant electron-phonon scattering. The observed value of Debye temperature $\theta_D$ from fitted resistivity plot is 151±0.1K, this value is verified through heat capacity measurements in the later part. The residual resistivity ratio was found to be 3.54, which is comparable to other topological single crystals [30,31]. Fig. 4(b) depicts magneto transport measurements of synthesized SnAs crystal performed at different fields viz. 0,20,40,60,100,120,140,160,180,200,250,300 and 500Oe. The superconducting transition temperature is gradually suppressed with the applied field. It is observed that superconductivity persists up to 250Oe, which is well above the critical field ($H_c$), being suggested in ref. 21. Transition width of superconductivity transition tends to increase as the applied field is increased, this can be attributed to vortex pinning in the sample. The value of $H_c$ at 0K i.e. $H_c(0)$, is determined by applying Werthamer-Helfand-Hohenberg (WHH) formalism, which is as follows

$$H_c(0) = -0.69 T_c \left[\frac{dH_c}{dT}\right]_{T_c} \quad (4)$$

The obtained value of $H_c(0)$ is 381Oe. Also, the obtained values of $H_c$ at different fields are plotted against temperature, and are fitted with the following quadratic equation known for conventional superconductors:

$$H_c = H_c(0) * \left[1 - \frac{T^2}{T_c^2}\right] \quad (5)$$

The fitted quadratic and WHH plots are shown in inset of Fig. 4(b). The values of $H_c$ at different temperature are found to be well fitted with quadratic equation, indicating SnAs to be a conventional superconductor. The obtained value of $H_c(0)$ from this formula is 360Oe, which is in agreement with the one as being calculated from WHH formalism. These values are also verified through heat capacity measurements in later part of this article.

Fig. 5(a) represents heat capacity measurements results (C/T vs T) of synthesized SnAs crystal in low temperature range of 2-6K at different magnetic viz. 0, 50, 100, 150, 200, 250 and 1000Oe. The superconducting transition temperature gradually decreases as the magnetic field is increased; the $T_c^{onset}$ is observed to be 3.7, 3.4, 3.1, 2.8, 2.5 and 2.2K at 0, 50, 100, 150, 200 & 250Oe. No transition is observed at 1000Oe. $T_c$ is found to have linear relationship with applied field, which is a trademark of conventional superconductivity. The inset of Fig. 5(a) shows the heat capacity measurements at zero field in temperature range 2-200K. As mentioned earlier, there are only two reports available on heat capacity measurements of SnAs [18,21]. Magnetic field dependent heat capacity is measured in ref. 18, and the heat capacity transition is observed up to 250Oe, which is certainly higher than the critical field suggested in ref. 21. Here also, the superconducting transition is visible up to 250Oe, which is in agreement with ref. 18. Also, it is observed that the heat capacity transition becomes broader as the applied magnetic field is increased. The magnitude of heat capacity jump is also found to be suppressed on application of magnetic field; this signifies that the sample enters in a mixed state. This behaviour is also observed in some other type-I superconductors [32]. Magneto heat capacity measurements along with the magneto transport measurements establish that the critical field for SnAs is above 250Oe. The value of upper critical field at absolute zero is calculated in later part by calculating the Sommerfeld coefficient.

Fig. 5(b) shows the C/T vs $T^2$ plot of synthesized SnAs crystal. The inset is showing the zero field C/T vs T plot in temperature range from 2 to 6K. A clear heat capacity jump is observed with $T_c^{onset}$ at 3.7K. This signifies the presence of bulk superconductivity in the synthesized crystal at 3.7K. Heat capacity of a material is a contribution of two terms at low temperatures viz. electronic contribution ($C_{el}$) and phonon contribution ($C_{ph}$). To determine both these contributions C/T vs $T^2$ plot is fitted with the following equation

$$\frac{C}{T} = \gamma_n + \beta_n * T^2 \qquad (6)$$

Here $\gamma_n T$ represents the electronic contribution to heat capacity, also $\gamma_n$ is known as Sommerfeld coefficient and $\beta_n T^3$ represents phonon contribution to heat capacity. The obtained value of $\gamma_n$ and $\beta_n$ are found to be 4.92 ± 0.26 mJ mol$^{-1}$ K$^{-2}$ and 0.87 ± 0.01 mJ mol$^{-1}$ K$^{-4}$ respectively. These values can be used to determine various normal state parameters of the synthesized SnAs crystal. Sommerfeld coefficient $\gamma_n$ is associated with Density of states (DOS) at Fermi level [$D_c(E_F)$] by following formula

$$\gamma_n = \frac{\pi^2 k_B^2 D_c(E_F)}{3} \qquad (7)$$

here $k_B$, represents Boltzmann constant and $D_c(E_F)$ is found to be 2.09 states eV$^{-1}$f.u.$^{-1}$. Debye temperature of the sample is related to the coefficient of phonon term by following formula

$$\theta_D = \left(\frac{12 \ ^4 nR}{5\beta_n}\right)^{1/3} \qquad (8)$$

here n is the number of atoms and R is ideal gas constant. For SnAs, the value of n is taken to be 2 and the calculated value of $\theta_D$ is found to be 161.6±0.08K. This value of $\theta_D$ is in agreement with the obtained value from resistivity data which is 151±0.1K. Now, the value of Debye temperature and critical temperature $T_c$ are used to calculate constant of electron phonon coupling $\lambda_{e-ph}$. $\lambda_{e-ph}$ is related to $T_c$ and $\theta_D$ through McMillan formula [33], given below

$$\lambda_{e-ph} = \frac{1.04 + \mu^* ln(\theta_D/1.45T_C)}{(1 - 0.62\,\mu^*)ln(\theta_D/1.45T_C) - 1.04} \tag{9}$$

here $\mu^*$ represents screened repulsive Coulomb potential. $\mu^*$ can take values from 0 to 0.2 for superconductors with $T_c$ between $10^{-3}$K to 20K as suggested in ref. 33, while in the same report, the empirical value of $\mu^*$ is suggested to be 0.13. Here, we took the empirical value of $\mu^*$ to be 0.13 as suggested in ref. 33. $T_c$ is taken as the temperature at the middle of the heat capacity jump. The obtained value of $\lambda_{e-ph}$ is 0.70, which is comparable to the values obtained for a weakly or intermediately coupled superconductor [33]. The electronic contribution to heat capacity ($C_{el}$) is determined by subtracting the phonon contribution term from the total heat capacity. $C_{el}$ is normalized by dividing $\gamma_n T_c$ and plotted against $T/T_c$, which is shown in Fig. 5(c). The magnitude of specific heat jump i.e., $\Delta C_{el}/\gamma_n T_c$ is found to be 1.41. This value is certainly below the cutoff value for BCS weakly coupled superconductors, which is 1.43. This value is in agreement with ref. 21. The value of heat capacity jump is used to determine and verify the upper critical field at absolute zero obtained from WHH formalism. Sommerfeld coefficient and heat capacity jump are related to critical field through following formula [32,34],

$$\triangle C = \frac{4H_c(0)^2}{\mu_0 T_c} = 1.43\gamma_n T_c \tag{10}$$

here, $\Delta C$ and $\gamma_n$ are taken in per unit volume. The molar volume of SnAs is $2.78\times10^{-5}m^3$/mole. Here heat capacity jump, $\frac{\Delta C}{\gamma_n T_c}$ is taken to be 1.41. The obtained value of $H_c(0)$ is found to be 333Oe, which is nearly equal to the critical field obtained from WHH formalism which is 360Oe. It is clear that the value of critical field is certainly higher than the reported value in ref. 21.

Conventional or unconventional nature of superconductivity can be determined through normalized electronic specific heat by calculating the value of parameter; $\alpha = 2\Delta(0)/k_B T_C$. To calculate $\alpha$, the normalized heat capacity data is fitted with the s wave superconductivity equation, and is shown in Fig. 5(c). This suggests that SnAs is a bulk superconductor with conventional s-wave pairing. The obtained value of α from the fitted plot is 3.36, which is also lower than the BCS limit for weakly coupled superconductors. The obtained value of α, corresponds to the superconducting energy gap $\Delta(0)$ of 0.54eV, which is comparable to that is obtained from Scanning tunneling spectroscopic (STS) measurements on the same sample [22] and as in ref. 18. As mentioned before, the only two reports available on heat capacity measurements of SnAs do contradict each other [18,21]. SnAs is shown to have BCS type weakly coupled superconductivity on the basis of heat capacity jump, which was observed below 1.43, while the value of α and superconducting gap were not calculated, which would have confirmed the statement. In ref. 18, the value of α is calculated by fitting the heat capacity data, and was found to be 3.73, i.e., above BCS weak coupling limit, indicating towards moderately coupled superconductivity in SnAs. Here, in our study all calculated parameters viz. α, $\lambda_{e-ph}$, $\Delta(0)$ and $\Delta C_{el}/\gamma T_c$ are found to lie in BCS weak coupling limit, confirming SnAs to be a weakly coupled s wave superconductor, which is in agreement with ref.21.

In Fig. 5(d), $C_{el}/T$ is plotted against T for different applied fields, and are fitted with the following equation [35] to calculate $\gamma_{res}$,

$$\frac{C_{el}}{T} = \gamma_{res} + A * \exp\left[\frac{-\Delta}{T}\right] \quad (11)$$

The obtained values of $\gamma_{res}$ are further plotted against different applied fields. The variation of $\gamma_{res}$ with respect to applied field, yields an important information about the low energy excitations, which take place very near to Abrikosov vortex line. These low energy excitations determine, whether the superconductor has conventional isotropic superconducting gap or it has a nodal superconducting gap. For a superconductor with an isotropic superconducting gap, the low energy excitations take place inside the vortex cores of having their radius proportional to penetration depth (ξ). For this, specific heat in superconducting state is proportional to vortex density and linearly depends on magnetic field giving $\gamma_{res}(H) \propto H$ [36]. While, in case of superconductors having nodal superconductivity, the DOS are found in the neighbourhood of gap nodes. This results in occurrence of the low energy excitations outside the vortex core and the specific heat is found to show square root dependence on magnetic field given as $\gamma_{res}(H) \propto H^{1/2}$, this is known as Volovik effect [37]. In the present case, $\gamma_{res}(H)$ is found to have a linear relationship with the applied field as shown in Fig. 5(e). This also confirms that the observed bulk superconductivity in synthesized SnAs crystal is of conventional nature with an isotropic superconducting gap. All the parameters obtained from heat capacity measurements are shown in Table-3.

In our previous report, which was based on magnetization measurements, the SnAs was found to be a weak type-II superconductor. Interestingly, here also, superconductivity is found to persist well above type-I critical field, as suggested in ref. 21. To clear the discrepancy in nature of superconductivity in SnAs, here another exercise is made, which includes calculation of superconducting critical parameters at T=0 viz. $\lambda(0)$ and $\xi(0)$. The unit cell of SnAs contains 4 formula units, while each formula unit provide three conduction electrons, providing in total 12 electrons per unit cell. Electron density is calculated by the ratio of number of electrons and volume of one unit cell i.e., n=12/V, where V= 187.25Å$^3$ (from Rietveld refinement). Electron density (n) is found to be 6.4×10$^{28}$ m$^{-3}$. Fermi wave vector is calculated by considering a spherical Fermi surface using the formula $k_F=(3n\pi^2)^{1/3}$, and is found to be 1.24Å$^{-1}$. Effective mass (m$^*$) is calculated by taking the ratio of Sommerfeld coefficient obtained from heat capacity measurement ($\gamma_n$) and the same as being obtained from theoretical calculations ($\gamma_{cal}$) in ref. 19. DOS at fermi level in theoretical calculations in ref. 19 is found to be approximately 0.80 states eV$^{-1}$f.u.$^{-1}$, providing $\gamma_{cal}$ to be 2.34 mJ mol$^{-1}$ K$^{-2}$. The ratio of $\gamma_n$ and $\gamma_{cal}$ is given by $\frac{\gamma_n}{\gamma_{cal}} = \frac{m^*}{m_e}$, providing m$^*$ to be 2.10m$_e$, where m$_e$ is free electron mass. The value of London penetration depth $\lambda(0)$ can be calculated by using the relation $\lambda(0) = \left(\frac{m^*}{\mu_0 n e^2}\right)^{1/2}$. The obtained value of $\lambda(0)$ is 30.4nm. The BCS coherence length is calculated using the formula $\xi(0) = \frac{0.18\hbar^2 k_F}{k_B T_C m^*}$, and it is found to be 245.5nm. Ginzberg Landau (G-L) κ parameter for a superconducting sample is given by the ratio of both the characteristic lengths viz. London penetration depth and BCS coherence length i.e., $\frac{\lambda(0)}{\xi(0)}$. κ parameter is found to be 0.12, which is certainly below the limit for type-I superconductivity which is $\kappa = \frac{1}{\sqrt{2}}$, suggesting SnAs to be a type-I superconductor. The obtained value of κ parameter is verified by calculating the Mean free path (l), using the formula $l = v_F \tau$, here v$_F$ is Fermi velocity and given by $v_F = \frac{\hbar k_F}{m^*}$, another term τ is scattering time and given by $\tau = \frac{m^*}{ne^2\rho}$, here ρ is taken as residual resistivity ρ=1.28×10$^{-6}$Ω-cm. The value of mean free path (l) is found to be 130.2nm. Certainly, $\xi(0)$ is very large as compared to the mean free path, suggesting SnAs to be a dirty limit superconductor. For a dirty limit superconductor, κ parameter can be calculated as $\kappa = \frac{0.75\lambda(0)}{l}$, the obtained value of κ is 0.17, which is nearly equal to the obtained value from the ratio of characteristic lengths, confirming the SnAs to be a type-I superconductor. This result agrees with ref. 18 and ref. 21. Our previous magnetization result indicating towards weak type-II superconductivity in SnAs [19] can be attributed to effect of sample shape, grain boundaries and domain walls as proposed in ref. 18. There have been some previous reports [38,39] on type-I superconducting single crystals, showing similar hysteresis loop in isothermal M-H curve as in ref. 19. Geometrical shape of the sample sometimes results in an un-accounted demagnetization factor, which results in such hysteresis in M-H loop [40]. Also, the strong pinning of domain walls is also suggested to be the reason for occurrence of hysteresis loop in M-H plot of type-I superconducting single crystals, as the same results again for an un-accounted demagnetization factor [39], similar to the geometrical shape effect. However, critical field at absolute zero in both magneto transport and magneto heat capacity measurements, is found to be well above than the previously reported value for type-I SnAs superconductor [18], with some glimpse of intermediate state. Still the nature of superconductivity in SnAs is unclear, here we warrant muon-spin rotation (μ-SR) spectroscopy measurements to be performed on SnAs crystals to calculate various characteristics lengths to clarify whether the same is a type-I or type-II superconductor.

**Conclusion:**

In this article, thermodynamic measurements are carried out to probe the nature of bulk superconductivity in the synthesized SnAs crystal. Bulk superconductivity is confirmed through magneto transport, AC susceptibility and heat capacity measurements. Some glimpses of intermediate state are observed in heat capacity measurements and a relatively high critical field is obtained in both magneto transport and heat capacity measurements. Also, the observed superconductivity in SnAs is found to be type-I, which is determined by calculating various superconductivity characteristics lengths. The heat capacity jump at the superconducting transition indicates SnAs to be a weakly coupled BCS superconductor. The value of parameter α is found to be 3.36, which also confirms that SnAs is a weakly coupled superconductor. Isotropic nature of superconducting gap is evident from linear dependency of $\gamma_{res}$ on applied field. Our report, establishes the fact that SnAs is a weakly coupled conventional BCS superconductor with an isotropic superconducting gap.

**Acknowledgement:**

Authors would like to thank Director NPL for his keen interest and encouragement. Authors would like to Ms. Prachi, IIT Roorkee, for XPS measurements. M.M. Sharma would like to thank CSIR-India for research grant and AcSIR for Ph.D. registration.

**Author's Statement:**

Authors do not have any conflict of interest.

## Table-1

Parameters obtained from Rietveld refinement:

| Cell Parameters | Refinement Parameters |
|---|---|
| Cell type: Face Centred Cubic (FCC) | $\chi^2$=3.51 |
| Space Group: F m -3 m | $R_p$=10 |
| Lattice parameters: a=b=c=5.721(0) Å | $R_{wp}$=13.3 |
| Cell volume: 187.255 Å$^3$ | $R_{exp}$=7.10 |
| Density: 6.961 g/cm$^3$ | |

## Table-2

| Element | Spin-orbit doublet | Binding Energy | FWHM |
|---|---|---|---|
| Sn | $3d_{5/2}$ | 484.20±0.01eV | 0.97±0.03eV |
| | $3d_{3/2}$ | 492.61±0.01eV | 0.75±0.04eV |
| As | $3d_{5/2}$ | 39.95±0.03eV | 0.82±0.09eV |
| | $3d_{3/2}$ | 40.65±0.03eV | 1.21±0.06eV |

## Table-3

Parameters obtained from Heat capacity measurements:

| Parameter | Obtained Value |
|---|---|
| $T_c^{onset}$ | 3.7K |
| $\gamma_n$ | 4.92 ± 0.26 mJ mol$^{-1}$ K$^{-2}$ |
| $\beta_n$ | 0.87 ± 0.01 mJ mol$^{-1}$ K$^{-4}$ |
| $D_c(E_F)$ | 2.01 |
| $\theta_D$ | 161.6±0.08K |
| $\lambda_{e-ph}$ | 0.70 |
| $\Delta C_{el}/\gamma_n T_c$ | 1.41 |
| $\alpha = 2\Delta(0)/k_B T_C$ | 3.36 |
| $\Delta(0)$ | 0.54meV |

**Figure Captions:**

**Fig. 1:** Rietveld refined Powder XRD pattern of SnAs in which the inset is showing the VESTA drawn unit cell of the same.

**Fig. 2:** (a) XPS spectra of SnAs in Sn 3d energy region (b) XPS spectra of SnAs in As 3d energy region.

**Fig. 3:** AC susceptibility measurements at 0 DC background of synthesized SnAs crystal.

**Fig. 4(a):** R-T plot at zero field of SnAs in which lower inset in showing the zoomed view of R-T plot in the proximity of superconducting transition while the upper inset is showing the fitted R-T plot in temperature range 2-90K.

**Fig. 4(b):** R-T plots of SnAs at different applied magnetic field from 0-500Oe in which inset is showing the variation of $H_c$ with respect to the temperature which is further fitted with WHH formalism and quadratic equation.

**Fig. 5(a):** Heat capacity measurements at different applied magnetic field in which inset is showing the same at zero field in temperature from 2K to 200K.

**Fig. 5(b):** Fitted C/T vs $T^2$ plot of synthesized SnAs crystal in which the inset is showing C/T vs T showing a clear superconducting transition with $T_c^{onset}$ at 3.7K.

**Fig. 5(c):** Normalized heat capacity vs $T/T_c$ fitted with conventional s wave pairing equation.

**Fig. 5(d):** $C_{el}/T$ vs T plot at different applied magnetic fields.

**Fig. 5(e):** Variation of $\gamma_{res}$ with respect to applied magnetic field fitted with linear equation.

Fig. 1

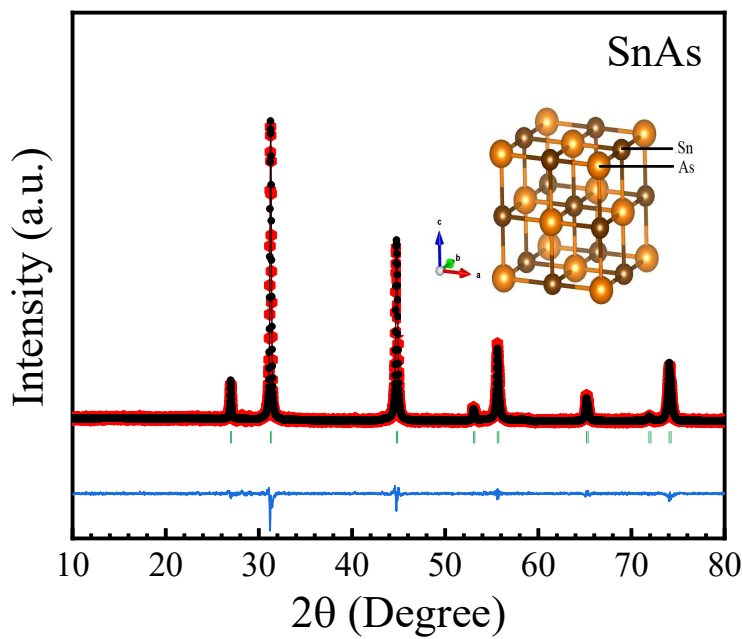

Fig. 2(a)&(b)

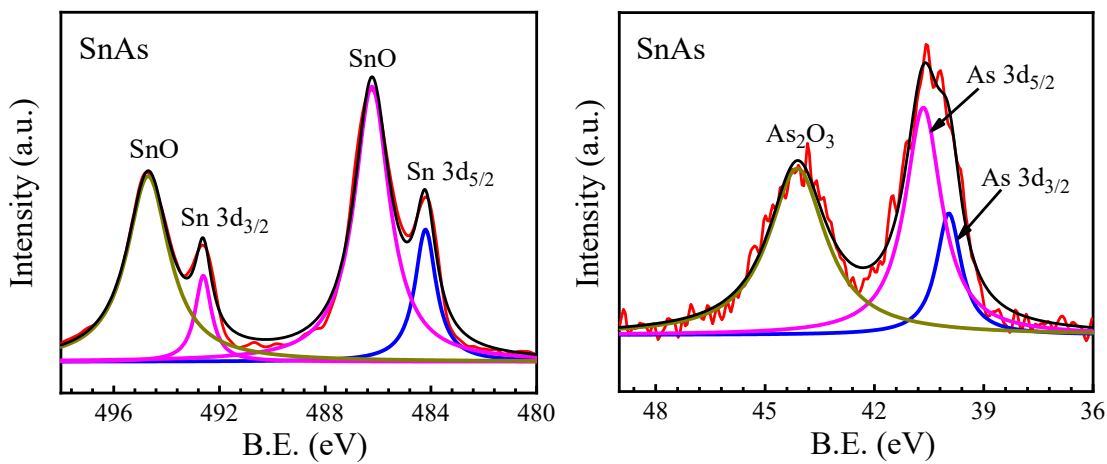

Fig. 3

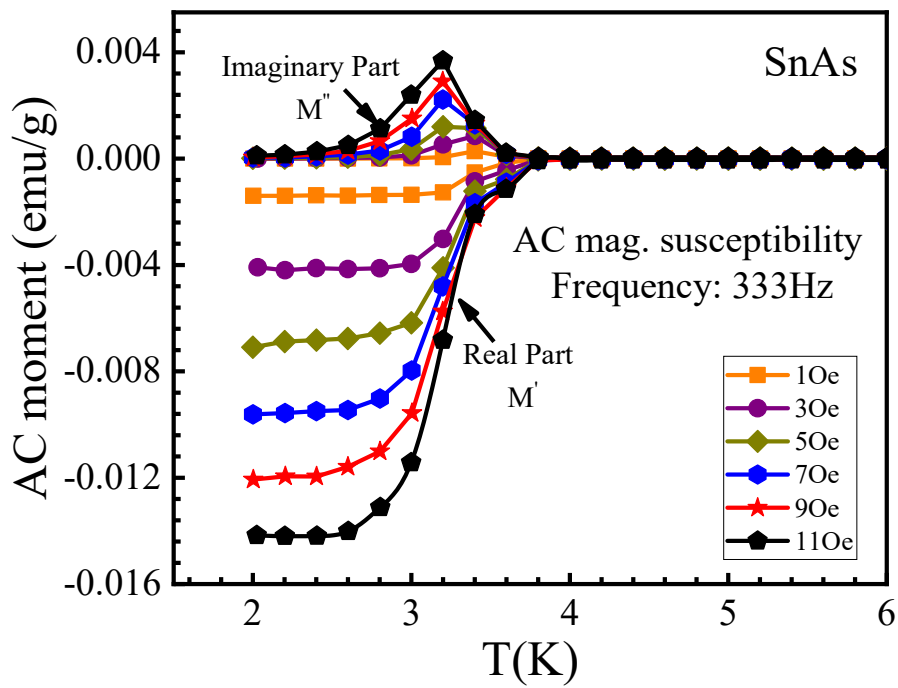

Fig. 4(a)

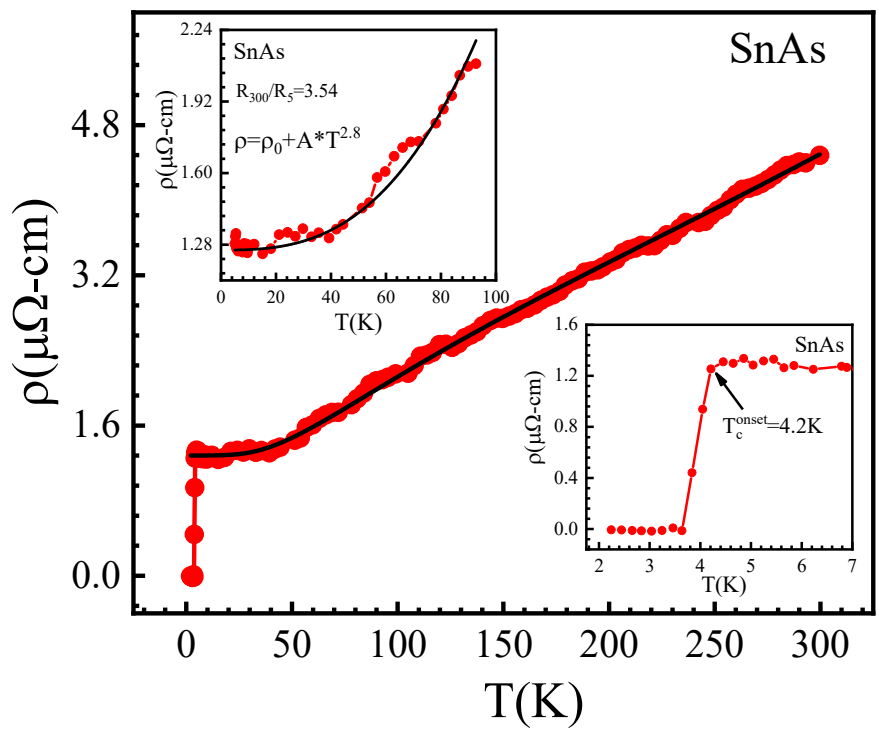

Fig. 4(b)

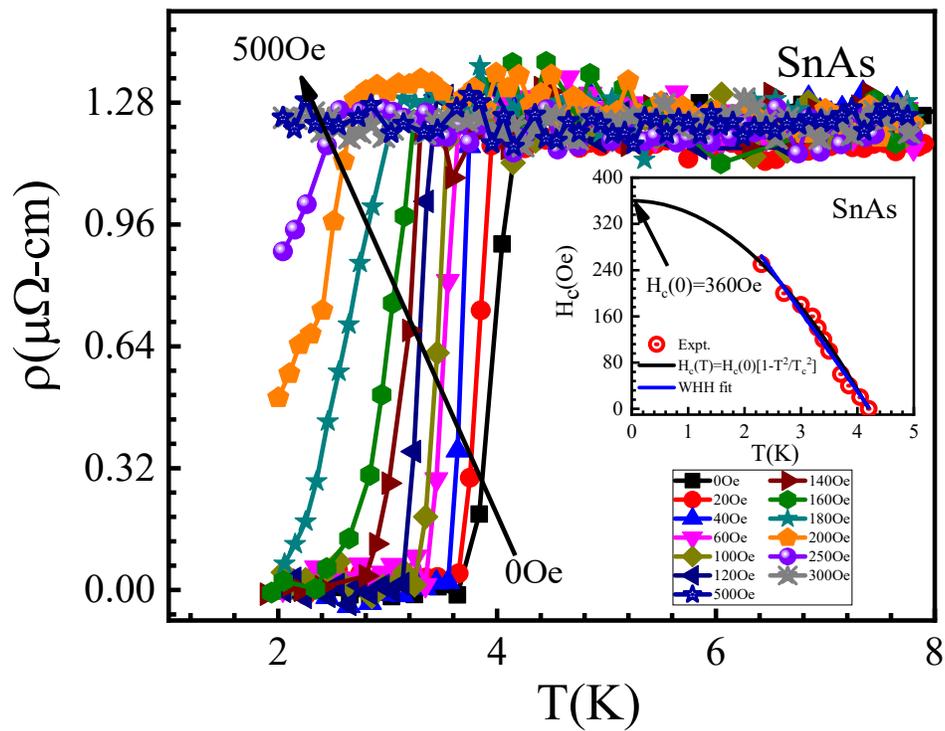

Fig. 5(a)

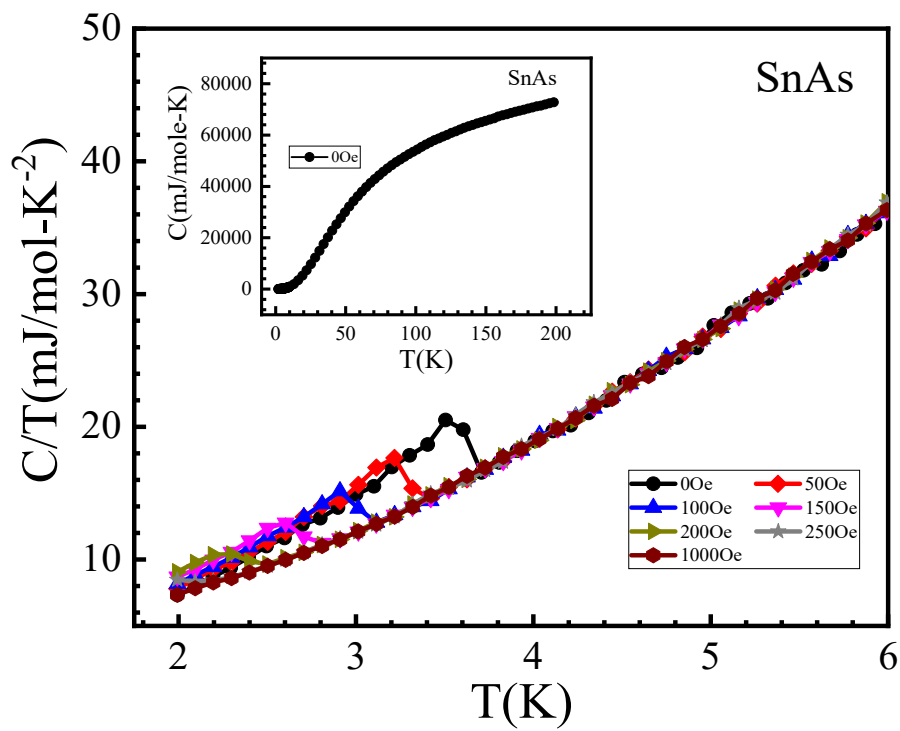

Fig. 5(b)

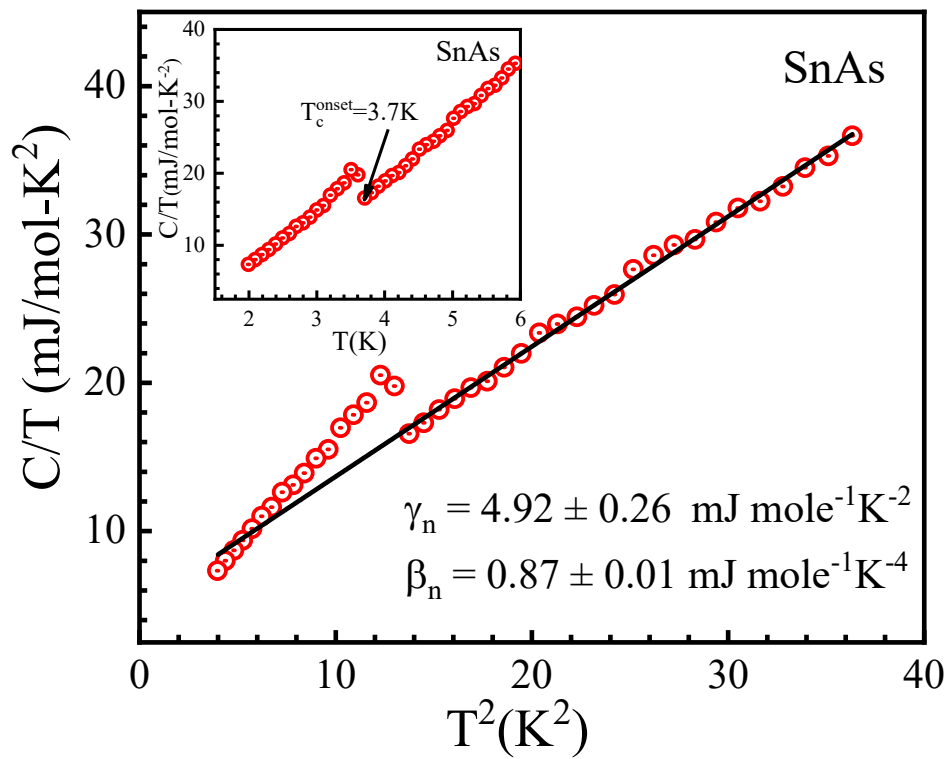

Fig. 5(c)

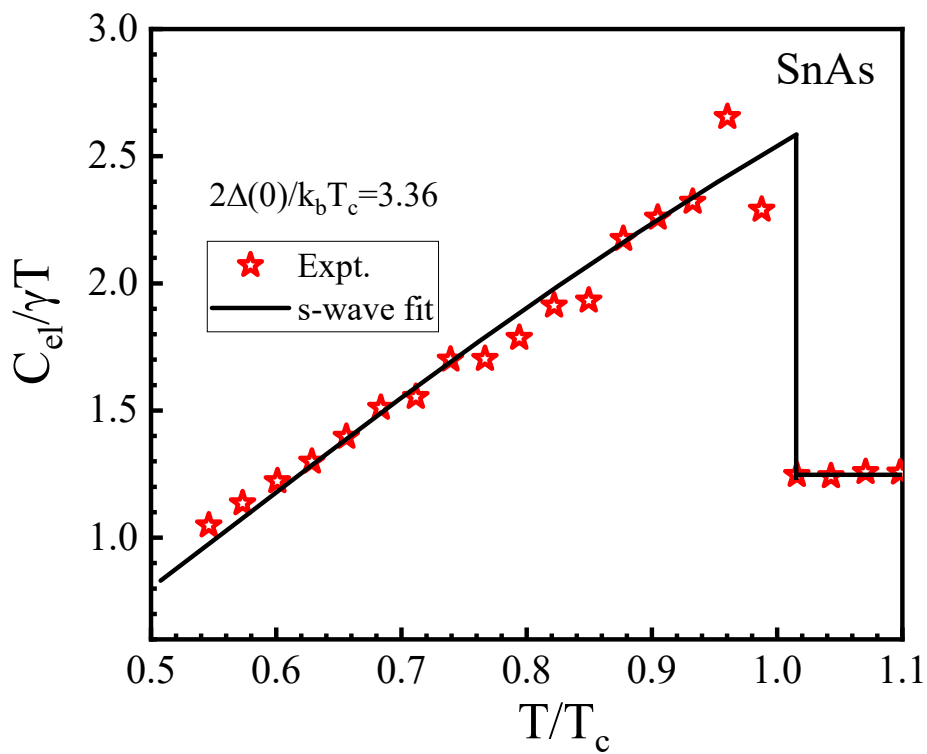

Fig. 5(d)

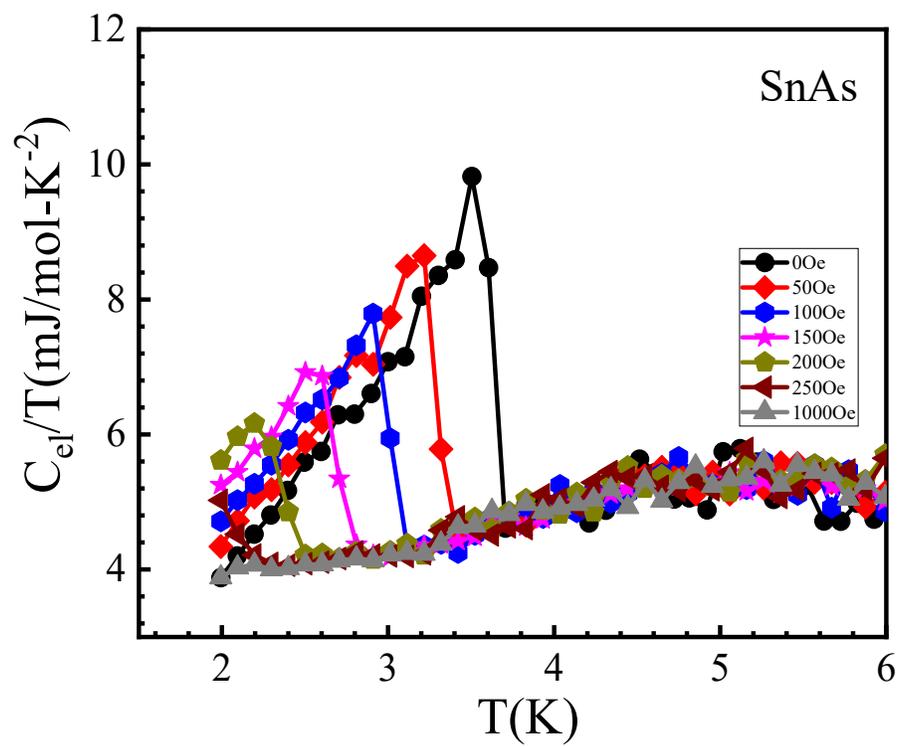

Fig. 5(e)

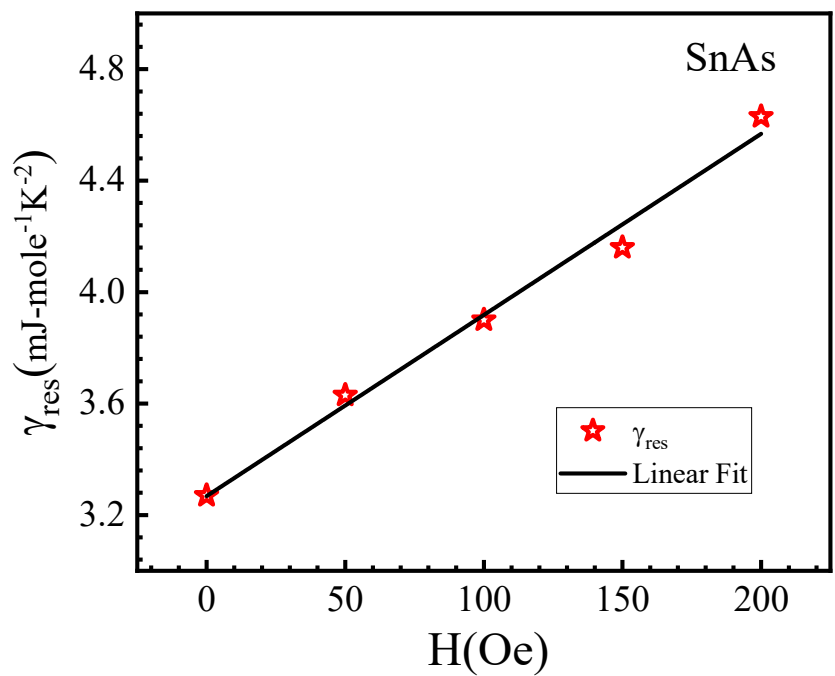